\documentclass[aps,pre,twocolumn,showpacs,superscriptaddress,groupedaddress]{revtex4}  
\usepackage{graphicx}  
\usepackage{dcolumn}   
\usepackage{bm}        
\usepackage{amssymb}   
\usepackage{amsmath}
\usepackage{xcolor}
\usepackage{float}

\begin{document}



\title{Controlling Escape in the Standard Map}

\author{Gabriel I. D\'iaz}
\affiliation{Instituto de F\'isica, IFUSP - Universidade de S\~ao Paulo, Rua do Mat\~ao, Tr.R 187, Cidade Universit\'aria, 05314-970, S\~ao Paulo, SP, Brazil}
\author{Matheus S. Palmero}
\affiliation{Instituto de F\'isica, IFUSP - Universidade de S\~ao Paulo, Rua do Mat\~ao, Tr.R 187, Cidade Universit\'aria, 05314-970, S\~ao Paulo, SP, Brazil}
\author{Iber\^e Luiz Caldas}
\affiliation{Instituto de F\'isica, IFUSP - Universidade de S\~ao Paulo, Rua do Mat\~ao, Tr.R 187, Cidade Universit\'aria, 05314-970, S\~ao Paulo, SP, Brazil}
\author{Edson D. Leonel}
\affiliation{Departamento de F\'isica, UNESP - Univ Estadual Paulista, Av. 24A, 1515, Bela Vista, 13506-900, Rio Claro, SP, Brazil}

\date{\today}

\begin{abstract}
We investigate how the diffusion exponent is affected by controlling small domains in the phase space.The main Kolomogorov-Arnold-Moser - KAM island of the Standard Map is considered to validate the investigation. The bifurcation scenario where the periodic island emits smaller resonance regions is considered and we show how closing paths escape from the island shore by controlling points and hence making the diffusion exponent smaller. We notice the bigger controlled area the smaller the diffusion exponent. We show that controlling around the hyperbolic points associated to the bifurcation is better than a random control to reduce the diffusion exponent. The recurrence plot shows us channels of escape and a control applied there reduces the diffusion exponent.

\end{abstract}


\maketitle

\section{\label{sec1}Introduction}

It is known that stickiness plays an important role regarding transport properties in several areas of physics, as fluids \cite{Fluids1,Fluids2}, plasma dynamics \cite{Plasma1,Plasma2} and celestial mechanics \cite{Celestial}. For the generic Kolmogorov-Arnold-Moser (KAM) scenario \cite{Kolmogorov1954,Moser1962,Arnold1963,Percival1979}, strong fluctuations are observed due to the presence of Cantori \cite{Meiss1992} acting as a partial barrier to the transport of particles. A way to characterize the stickiness phenomena is by using the diffusion exponent \cite{Scafetta2002,Diaz2019}, whose value allows us to distinguish between anomalous diffusion, when the exponent is different from $1/2$, and normal diffusion, when the exponent is $1/2$.  

In a previous work \cite{Diaz2019}, we applied a method to calculate the diffusion exponent of an initial ensemble of orbits around  Kolmogorov-Arnold-Mose islands. We showed the diffusion exponent changed when the island experienced a bifurcation generating resonance islands. Based in the work of \cite{Sun2005,Kruger2015,Contopoulos2010} we conjectured that this happened by the manifold dynamics of the hyperbolic fixed points associated to the resonance islands. In this work we investigate deeper such a conjecture. We control small areas in phase space around the KAM islands, where every time that a orbit enters these areas it is re-initiated as a random point of the initial ensemble. Delimiting the escape of the orbits by that zone in phase space we are able to change the diffusion exponent and determine if around the hyperbolic point exists channels of escape as stated in \cite{Diaz2019}.

The paper is organized as follows, section \ref{sec2} introduces the model under study and briefly describes the method to calculate the diffusion exponent \footnote{More details can be found in \cite{Diaz2019}.}, in section \ref{sec3} we present some previous results, in section \ref{sec4} we show our numerical results and finally conclusions are drawn in section \ref{sec5}.

\section{\label{sec2} Description of the model and the method}

\begin{figure}[h!]
\begin{centering}
\includegraphics[angle=-90,scale=0.3]{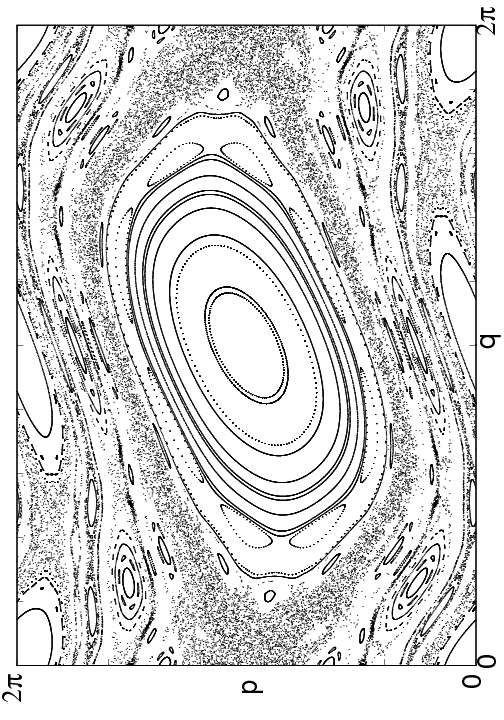}
\par\end{centering}
\caption{Plot of the phase space for the Standard Map considering the control 
parameter $k=1$. \label{fig:stdmapk1}}
\end{figure}

The model under study is the Standard Map \cite{Chirikov1969,Chirikov1979}, which describes the motion of a particle constrained to move in a ring. The particle is kicked periodically by an external field. The dynamics of the Standard Map (SM) is described by a mapping $T_{SM}\left(p_{n},q_{n}\right)=\left(p_{n+1},q_{n+1}\right)$ that gives the position and momentum just before the $(n+1)^{th}$ kick
\begin{equation}
T_{SM}:\left\{\begin{array}{ll}
p_{n+1} =[p_n +k\sin(q_n)]\mod(2\pi)\\
q_{n+1} = [q_n + p_{n+1}]\mod(2\pi)\\
\end{array}
\right.,
\label{eq:SM}
\end{equation}
where the parameter $k$ controls the intensity of the non-linearity of the mapping. Since the determinant of the Jacobian matrix is the unity, the mapping preserves the area on the phase space. A plot of the phase space of SM is shown in Fig. \ref{fig:stdmapk1}.

One sees from Fig. \ref{fig:stdmapk1} a coexistence of chaotic domains around regular ones, the regions of regular motion are generally formed by invariant curves arranged in complex structures called Kolmogorov-Arnold-Moser (KAM) islands \cite{Kolmogorov1954,Moser1962,Arnold1963,Percival1979}.

We follow the procedure presented in \cite{Diaz2019} to measure the diffusion exponent around KAM islands. The procedure is based in the studies of Scafetta and Grigolini \cite{Scafetta2002} to determinate the diffusion exponent by using entropy measurements. A Brief description of the method is the following, (for more details see \cite{Diaz2019}):

\begin{enumerate}
    \item Choose a two dimensional $I$ (horizontal divisions) $\times I$ (vertical divisions) grid with help of the relation $10\times\frac{diffusion\;\; range}{fluctuation\;\;size} \sim grid\;\;divisions$.
    \item Set an ensemble of initial conditions around the KAM island to apply the mapping.
    \item Apply the mapping \ref{eq:SM} to the ensemble of orbits. At each iteration construct an histogram $[h_{ij}]$ counting how many points are inside each grid box. Then measure the entropy by means of the equation $S = -\sum_{i=1}^{I}\sum_{j=1}^{I}h_{ij}\ln\left(h_{ij}\right)$.
    \item After some iterations search of a interval in time where the entropy grows linearly\footnote{In this region is possible to consider the scaling hypothesis of \cite{Scafetta2002}.} with $\ln\left(n\right)$.
    \item Use the equation $S = A+\delta\ln\left(n\right)$ to calculate the diffusion exponent $\delta$.
\end{enumerate}

A variant of the method where we control an small area in phase space has to consider a conditional statement {\it if} in the SM. We consider additional steps in the previous method:

\begin{itemize}
    \item[$2'$]  Choose a control area in phase space \footnote{In our case an circular ball of some radius.}.
    \item[$3'$] At every iteration for each element of the ensemble of orbits ask if the orbit has
    entered in the control area. If it does it, re-initiate the orbit choosing a new initial condition randomly from our ensemble of orbits at $n = 0$. 
\end{itemize}

\section{\label{sec3} Previous results}

Similarly to \cite{Diaz2019} we measure the area of the main island in the SM and compare it 
to the diffusion exponent. To get the area of the main island 
separating chaos from regular behavior in the phase space we see what regions 
chaotic orbits visit and what regions are not visited. We consider values of the parameter $k$ where a set of chaotic orbits can visit all the chaotic sea provided enough time is allowed. We divide the phase space into grid of boxes and ask if any of the chaotic orbits visited a given grid element. If so we mark it
as yellow indicating a box of chaotic behavior. If none orbit has landed in a box we paint is as a black characterizing a box of regular behavior.

The main island’s center is an elliptic fixed point of coordinates $q = \pi$ , $p = \pi$. It is located inside a box of regular domain, hence black. Selecting all the boxes of black color that are connected, i.e., that are first nearest neighbors, starting from that one, we are able to find an approximation of the 
main island area by a black region of simple connected boxes. We call to the total area of this boxes divided by total area, $(2\pi)^2$, the normalized area.  

\begin{figure}[h!]
\begin{centering}
\includegraphics[angle=-90,scale=0.3]{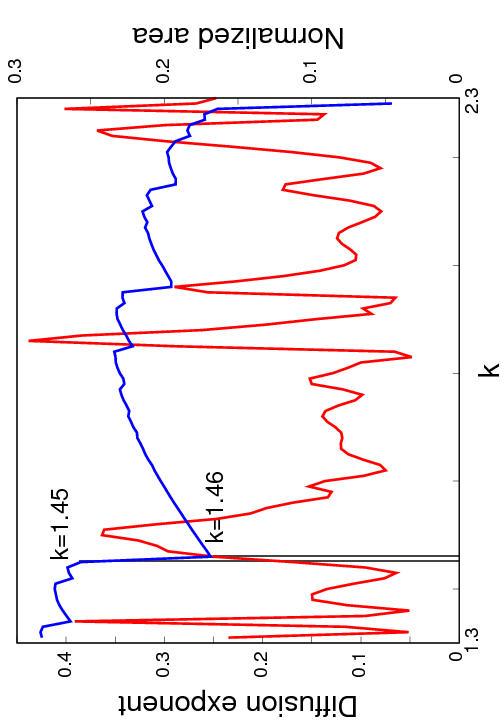}
\par\end{centering}

\caption{Plot of the diffusion exponent (red line) and the normalized area of 
the main island (blue line) vs. the parameter $k$. The region between vertical 
lines corresponds to the interval $k \in [1.45,1.46]$ (Colors on line).
\label{fig:dif_area_vs_k}}

\end{figure}

It is possible to see in Fig. \ref{fig:dif_area_vs_k} that whenever the area decreases abruptly the diffusion exponent increases. Furthermore the area grows, while the exponent decreases, until a critical value when the area abruptly decreases once more, with its corresponding increase in the diffusion exponent. We mark two values of $k$, region between the vertical lines in Fig. \ref{fig:dif_area_vs_k}, at this interval happens the largest decrease in the area, for the values of $k$ considered.

\begin{figure}[h!]
\begin{centering}

$k=1.45$
\includegraphics[angle=-90,scale=0.3]{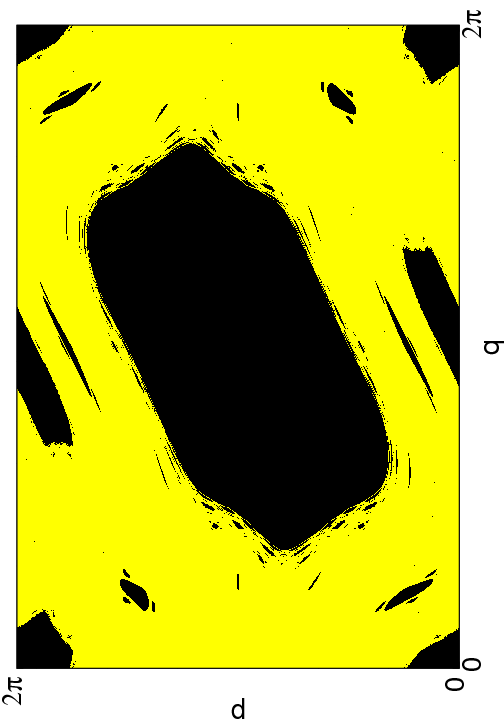}

~\\
~\\

$k=1.46$
\includegraphics[angle=-90,scale=0.3]{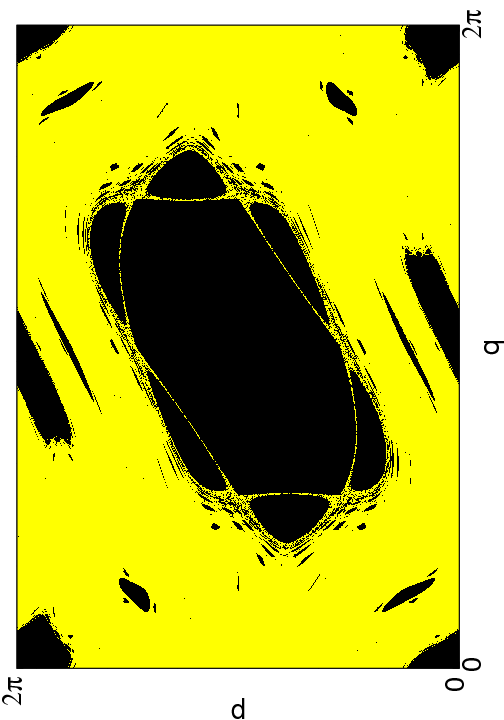}

\end{centering}
\caption{Figure showing the separation of chaotic sea marked in gray (yellow color) from the regular islands marked in black color for two values of the control parameter. Smaller islands are ejected from the main island when the control parameter changed (Colors on line). \label{fig:border_ka_stdm}}

\end{figure}

We see in Fig. \ref{fig:border_ka_stdm} the transition marked in Fig. \ref{fig:dif_area_vs_k}. When passing from $k=1.45$ to
$k=1.46$ the main island ejects a resonance of smaller islands, therefore reducing the area of the main island. However, each ejected island has an elliptic periodic point in the middle, and by the Poincar\'e Birkhoff theorem \cite{Lange2014,Birkhoff1925} 
exists their corresponding hyperbolic fixed points pair. According to the conjecture presented in Ref. \cite{Diaz2019} the action of the stable and unstable manifolds of such hyperbolic points is responsible for the changes in diffusion behavior since they provide large channels  to escape from the main island \cite{Sun2005,Kruger2015,Contopoulos2010}.

\section{\label{sec4} Numerical results}

For a given value of $k$ from the SM. Eq. \ref{eq:SM}, we search the hyperbolic points, associated to the smaller islands
ejected, and control a circular area around them \footnote{Where the orbits are reinitialized near the island when enter in the circular ball, as indicated in the variant method of Sec. \ref{sec2}.} when we calculate
the diffusion exponent.

\begin{table}[h!]
\begin{tabular}{|c|c|c|c|}
\hline 
$k$ & Period & $\delta$ (with control) & $\delta$ (no control)\tabularnewline
\hline 
$1.46$ & $6$ & $0.143\pm0.001$ & $0.247\pm0.001$\tabularnewline
\hline 
$1.47$ & $6$ & $0.2448\pm0.0009$ & $0.3218\pm0.0009$\tabularnewline
\hline 
$1.48$ & $6$ & $0.152\pm0.003$ & $0.288\pm0.003$\tabularnewline
\hline 
$1.85$ & $16$ & $0.355\pm0.002$ & $0.358\pm0.002$\tabularnewline
\hline 
$1.86$ & $16$ & $0.379\pm0.005$ & $0.383\pm0.005$\tabularnewline
\hline 
$1.87$ & $16$ & $0.397\pm0.002$ & $0.399\pm0.002$\tabularnewline
\hline 
$1.96$ & $10$ & $0.201\pm0.009$ & $0.221\pm0.009$\tabularnewline
\hline 
$1.97$ & $10$ & $0.25\pm0.01$ & $0.275\pm0.009$\tabularnewline
\hline 
$1.98$ & $10$ & $0.179\pm0.007$ & $0.189\pm0.007$\tabularnewline
\hline 
\end{tabular}
\caption{Table showing changes in the diffusion exponent when controlling an small circular
area around hyperbolic points, (ball radius $r=0.001$) \label{tab:pntcontr_chngdelt_kvarius}.}
\end{table}

In table \ref{tab:pntcontr_chngdelt_kvarius} we see the change in
diffusion exponent when the area around the hyperbolic points is controlled. For different values of the parameter $k$ we have different periods in the resonance islands, but in every case the diffusion exponent become smaller than without control. The values of $k$ were chosen near three decays in the normalized area from Fig. \ref{fig:dif_area_vs_k}. 

\begin{figure}[h!]
\begin{centering}
\includegraphics[scale=0.3]{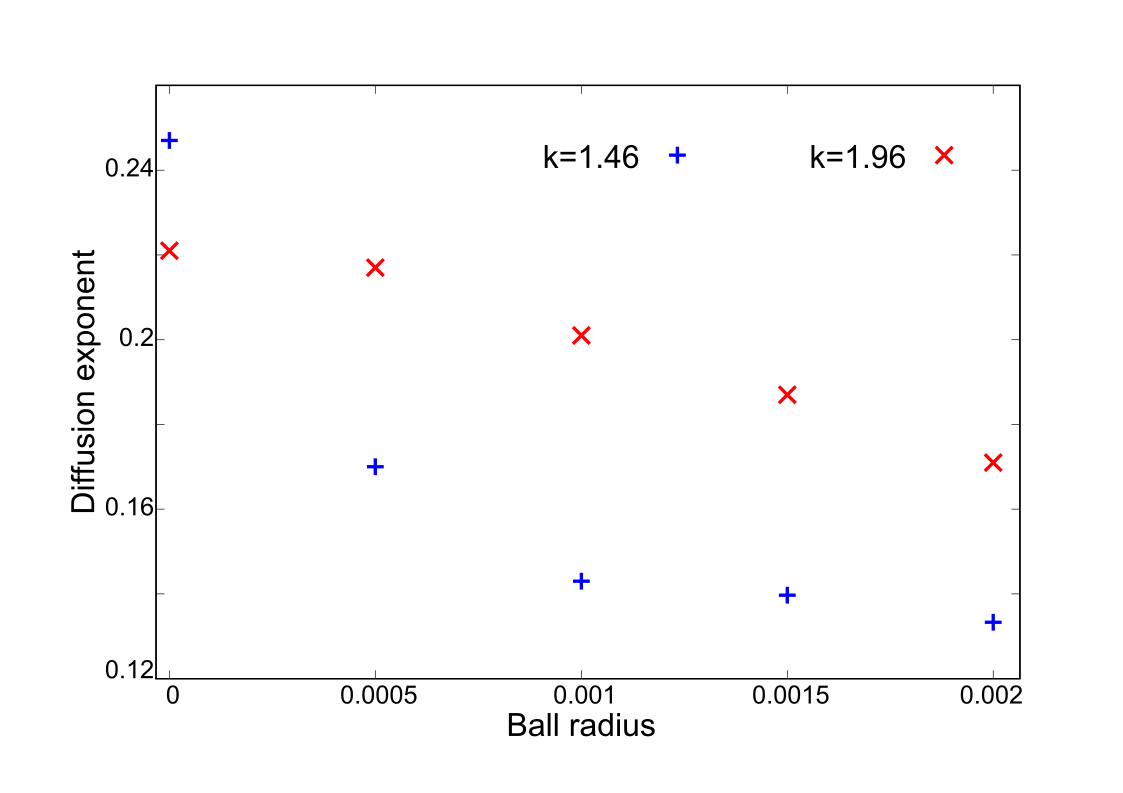}
\par\end{centering}

\caption{Plot of the diffusion exponent against the radius of the controlling ball for two values of the control parameter, namely $k=1.46$ and $k=1.96$.
\label{fig:radcontr_chngdelt_kvarius}}
\end{figure}

We examine now the effect of the control area, we change the radius of the circular ball around the hyperbolic points and see how does the diffusion exponent changes, see Fig. \ref{fig:radcontr_chngdelt_kvarius} 

It is possible to see in Fig. \ref{fig:radcontr_chngdelt_kvarius}
that a bigger value radius of control area translates into a smaller value of the diffusion exponent, meaning that more orbits are re-initiated since it is more likely than a orbit enters in the control area. If we change the position of the control area, the diffusion exponent also changes, as can be seen in Figs. \ref{fig:pntcontr_chngdelt_k1,46} and \ref{fig:pntcontr_chngdelt_k1,96}. We we notice the diffusion exponent seems to be more affected when the control area is around an hyperbolic point (red point) than when is around another random point (blue points).

\begin{figure}[h!]
\begin{centering}
\includegraphics[angle=-90,scale=0.3]{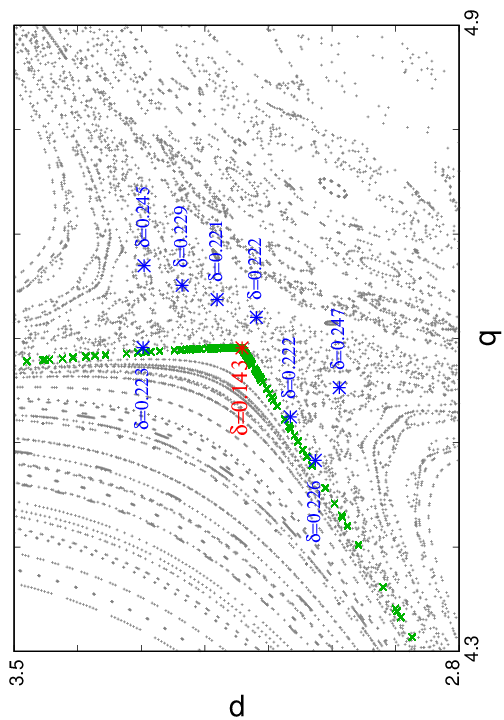}
\par\end{centering}

\caption{Plot of diffusion exponent for $k=1.46$, the position of the control point is indicated with blue points with its diffusion exponent next to them. The red point indicates the hyperbolic fixed point. The green points are the initial ensemble of conditions. The gray dots in the background shows the Standard Map portrait for that value of $k$ (Colors on line).
\label{fig:pntcontr_chngdelt_k1,46}}
\end{figure}

\begin{figure}[h!]
\begin{centering}
\includegraphics[angle=-90,scale=0.3]{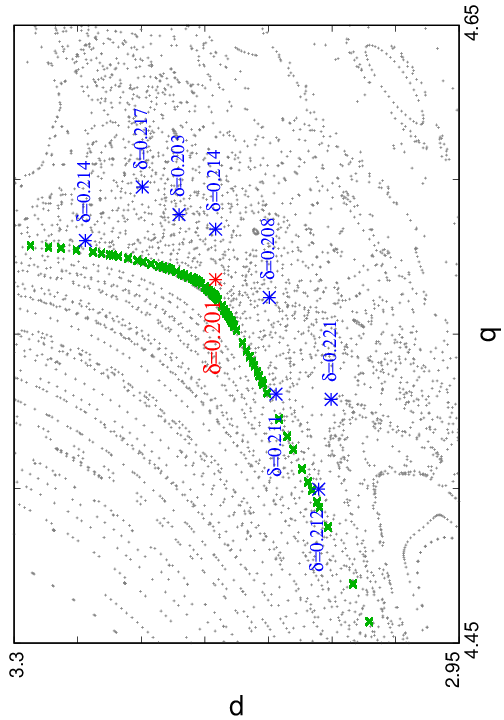}
\par\end{centering}

\caption{Plot of the diffusion exponent for $k=1.96$, the position of the control point is indicated with blue points with its diffusion exponent next to them. The red point indicates the hyperbolic fixed point. The green points are the initial ensemble of conditions. The gray dots in the background shows the Standard Map portrait for that value of $k$ (Colors on line).
\label{fig:pntcontr_chngdelt_k1,96}}
\end{figure}

Finally we search if there is a connection between the diffusion
exponent and the recurrence in phase space. For an initial ensemble
of conditions we calculate the number of times any of them enters a small box of a grid. According to the number of times any point entered, a box is assigned with a 
given color, see Fig. \ref{fig:recurrence_k1,46}.

\begin{figure}[h!]
\begin{centering}
\includegraphics[angle=-90,scale=0.35]{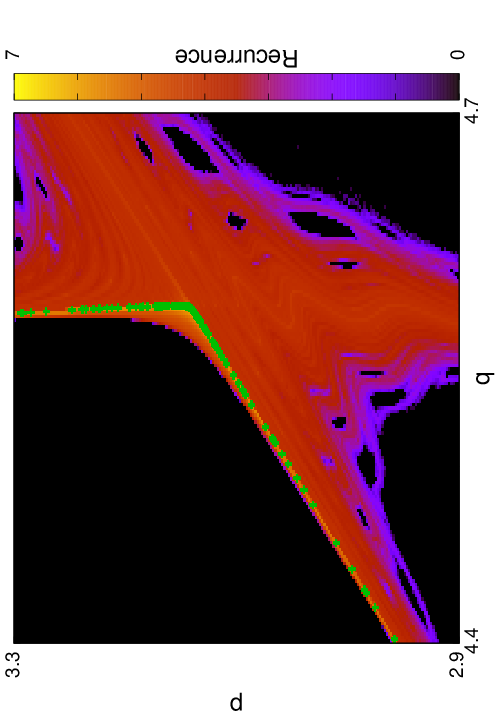}
\par\end{centering}

\caption{Recurrence plot for the initial ensemble of orbits (green points).
Every box in the grid has a color assigned depending on the number of points visited them, yellow indicates a larger concentration of  points, black a small concentration (Colors on line).
\label{fig:recurrence_k1,46}}
\end{figure}

\begin{figure}[h!]
\begin{centering}
\includegraphics[angle=-90,scale=0.35]{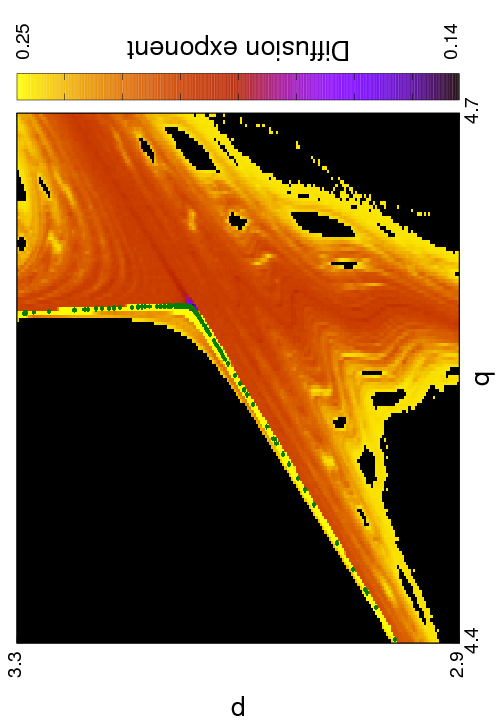}
\par\end{centering}

\caption{Diffusion exponent for the initial ensemble of orbits (green points).
Every box in the grid has a color assigned depending on the value of the diffusion exponent when the control area is in the grid box (Colors on line).
\label{fig:grid_expo_k1,46}}

\end{figure}

In Fig. \ref{fig:grid_expo_k1,46} wee see a color grid with the value of the diffusion exponent when the control area is in each grid box. We notice the diffusion exponent near the hyperbolic point it is smaller than in other regions. Even more, this figure has similar structure as shown Fig. \ref{fig:recurrence_k1,46} confirming small channels were the control is better than away from those channels.      

\section{\label{sec5} Conclusion}

We calculated the diffusion exponent around the main island in the Standard map. We focused in the bifurcation scenario were the main KAM island emits smaller resonance islands. For different values of the nonlinear parameter $k$ we showed that when considering an small control area, around hyperbolic points associated to the resonances, it is possible to change the diffusion exponent making it smaller. This is mainly due to the fact that the control action closes paths to escape from the main island. We also showed that the bigger the control area the smaller the diffusion exponent, since it is more probable that an orbit will enter the control area.

We showed that changing the position of the control point the diffusion exponent changes, for random control points this diffusion exponent is not much affected as when we consider the hyperbolic point.

Finally, when comparing Fig. \ref{fig:recurrence_k1,46}  and Fig. \ref{fig:grid_expo_k1,46}, we saw that a recurrence plot shows channels of escape, places of high recurrence, where a controlled point made the diffusion exponent smaller.   

\section*{ACKNOWLEDGMENTS}
G.I.D. thanks the fellowship from National Council for Scientific and Technological Development (CNPq). M.S.P., I.L.C., and E.D.L. acknowledge Sao Paulo Research Foundation (FAPESP) Grants No. 2018/03000-5, No. 2018/03211-6, and No. 2019/14038-6. I.L.C. and E.D.L. acknowledge National Council for Scientific and Technological Development (CNPq) Grants No. 300632/2010-0 and No. 301318/2019-0.   

\end{document}